\documentclass[amsmath,amssymb]{revtex4}
\newcommand {\eq}{\begin{equation}}
\newcommand {\qe}{\end{equation}}

\newcommand {\h}{\frac{1}{2}}

\newcommand {\np}{Nucl. Phys. }

\usepackage{graphicx}
\usepackage{epsfig}
\begin{document}

\title{Electromagnetic and Mass Difference Corrections in $\pi$ N 
Scattering}

\author{ W. R. Gibbs and R. Arceo}
\affiliation{ Department of Physics, New Mexico State University \\
 Las Cruces, New Mexico 88003, USA\\}

\begin{abstract}

Coulomb and mass difference corrections to low-energy pion-nucleon
scattering are calculated and compared with previous work including 
potential models, dispersion relation methods and chiral perturbation 
theory calculations. Particular attention is paid to their role in 
testing isospin breaking.

\end{abstract}

\maketitle

\section{Introduction}

Electromagnetic corrections are a key element in understanding low
energy pion nucleon scattering and, in turn, the fundamental elements
of chiral symmetry breaking and the pion-nucleon coupling constant.
They are also crucial in the important role of low-energy pion-nucleon 
scattering as a testing ground for isospin breaking.

Tests of isospin breaking have been reported\cite{gakl, matsinos,gak}
using the extracted charged pion scattering amplitudes to predict charge
exchange amplitudes. A breaking of the order of 8\% in the amplitude was
found. Piekarewicz\cite{jose} has suggested a possible explanation for the
isospin breaking based on the effect of quark mass differences on the
charged and neutral pion coupling constants.

The breaking mentioned above is beyond the expected corrections due to
electromagnetic interactions and mass differences. Corrections for these
effects have been reported using a potential model for their
calculation\cite{gashi1,gashi2,zimmermann} based on pioneering work by
Oades and Rache\cite{oades}.  In addition, an analysis was made\cite{gak}
(hereafter referred to as GAK)  in which the potential method corrections
were incorporated into the fit itself.  These last corrections have not
been previously reported and it is a primary purpose of this paper to
give them.

For unbroken isospin one can write the charged pion scattering amplitudes
in terms of the isospin amplitudes.

\eq
f_{\pi^+p\rightarrow\pi^+p}=f^{\frac{3}{2}};\ \ 
f_{\pi^-p\rightarrow\pi^-p}=\frac{1}{3}(f^{\frac{3}{2}}+2f^{\h})
;\ \ \ f_{\pi^-p\rightarrow\pi^0n}=\frac{\sqrt{2}}{3}(f^{\frac{3}{2}}
-f^{\h})=\frac{1}{\sqrt{2}}\left( f_{\pi^+p\rightarrow\pi^+p}-
f_{\pi^-p\rightarrow\pi^-p}\right)
\qe

In the potential treatment each of the pure isospin amplitudes is assumed
to satisfy a wave equation with potentials corresponding to isospin
($V^{\frac{3}{2}}$ and $V^{\frac{1}{2}}$).  Since the $\pi^+$p amplitude
consists of a single isospin amplitude, only the Coulomb correction needs
to be taken into account.  The method for this case is covered in Section
\ref{piplusp}.

The last two amplitudes, being a mixture of isospin amplitudes, pose the
additional problem of the neutron-proton and $\pi^{\pm}$--$\pi^0$  mass 
differences. By using a linear combination of the amplitudes, the solution 
is expressed as a pair of coupled equations corresponding to the charge 
states.

\newpage

\eq
{\cal 
T}_c\psi_c+\left(\frac{2}{3}V^{\frac{1}{2}}+\frac{1}{3}V^{\frac{3}{2}}
\right)\psi_c+
\frac{\sqrt{2}}{3}\left(V^{\frac{3}{2}}-V^{\frac{1}{2}}\right)\psi_0=
\epsilon_c\psi_c
\qe

\eq
{\cal T}_0\psi_0+
\frac{\sqrt{2}}{3}\left(V^{\frac{3}{2}}-V^{\frac{1}{2}}\right)\psi_c+
\left(\frac{1}{3}V^{\frac{1}{2}}+\frac{2}{3}V^{\frac{3}{2}}\right)\psi_0=
\epsilon_0\psi_0
\qe
where $\psi_c$ and $\epsilon_c$ are the charged wave function and kinetic
energy and $\psi_0$ and $\epsilon_0$ are the corresponding neutral pion
quantities. ${\cal T}_c$ and ${\cal T}_0$ are the kinetic energy 
operators. The corrections are then included by adding a Coulomb
potential to the equation describing the $\pi^-$ only.  The masses
entering into the kinetic energy operators, ${\cal T}$, and the kinetic
energies, $\epsilon$, are also replaced by expressions with the true 
masses corresponding to the charged or neutral pions and nucleons.  It is 
assumed that the potentials remain unchanged during this process.  This 
assumption has been questioned by Rusetsky\cite{rusetsky}.

Although the general method used in each case is very similar, each group
has used different forms for the potentials, fit different data sets, used
different relativistic prescriptions for the pion energies and reduced
masses (energies) and, in some cases, different wave equations.  Thus,
variations are to be expected among the resulting corrections.

In addition to the potential method, a dispersion approach has been used
\cite{nordita,norditab}. Sauter\cite{sauter} developed the basic method
for s-waves only without considering mass differences. The model was
extended in several works\cite{disp}.  This approach has become the most
common method for making the corrections and they are  often referred to 
as 
the NORDITA\cite{nordita,norditab} results. The advantages and 
disadvantages of the potential and dispersion relations method are 
discussed in Ref. \cite{gashi1}.

 \section{$\pi^+$ proton scattering\label{piplusp}}

In this case the corrections in a potential model are straightforward.  
The wave equation can be solved for each partial wave to obtain the
phase shifts. While this full solution for the amplitude includes 
everything, the partial 
wave expansion diverges, as does the one for a pure Coulomb potential.   
Since the sum of the Coulomb series alone is known, the non-spin-flip and 
spin-flip amplitudes can be written as
\eq
f(\theta)=f_c(\theta)+\frac{1}{2ik}\sum_{\ell=0}^{\infty}
[(\ell+1)(e^{2i\delta_{\ell+}}-e^{2i\sigma_{\ell}})
+\ell(e^{2i\delta_{\ell-}}-e^{2i\sigma_{\ell}})]P_{\ell}(\cos\theta)
\label{basic}\qe
\eq
g(\theta)=\frac{1}{2k}\sum_{\ell=0}^{\infty}[e^{2i\delta_{\ell+}}
-e^{2i\delta_{\ell-}}]P_{\ell}^1(\cos\theta)
\qe
where $\sigma_{\ell}$ is the Coulomb phase shift and $f_c(\theta)$
is the Coulomb amplitude. The notation $\ell\pm$ means $j=\ell\pm\h$.
Here (and in GAK\cite{gak}) (unlike Gashi et al.\cite{gashi1}) we neglect 
the spin dependent part of the electromagnetic interaction.
Eq. \ref{basic} is often written as
\eq
f(\theta)=f_c(\theta)+\frac{1}{2ik}\sum_{\ell=0}^{\infty}
e^{2i\sigma_{\ell}}[(\ell+1)(e^{2i\delta^n_{\ell+}}-1)
+\ell(e^{2i\delta^n_{\ell-}}-1)]P_{\ell}(\cos\theta)
\qe
where the nuclear phase shift is defined as
\eq
\delta^n_{\ell\pm}=\delta_{\ell\pm}-\sigma_{\ell}.
\qe

\begin{center}
\begin{figure}[htb]
\epsfig{file=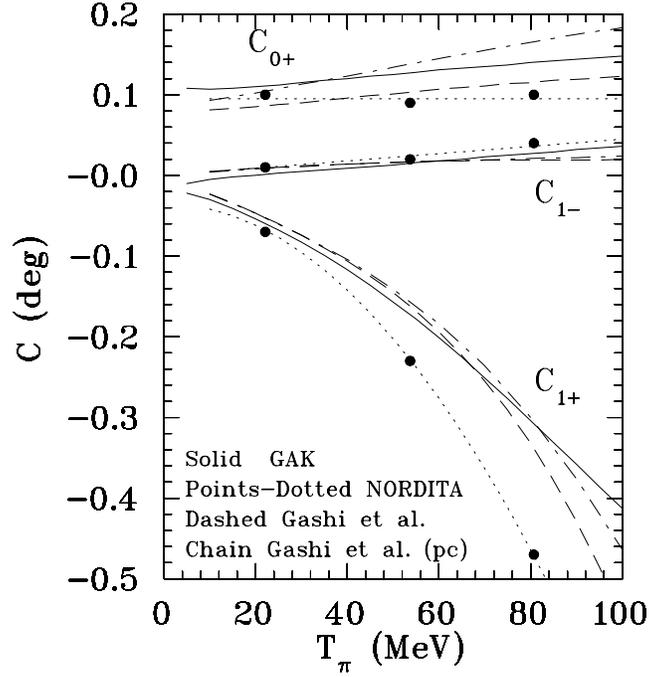,height=3.5in} 
\caption{Comparison among the different method for $\pi^+$p elastic
scattering. The difference between the point charge (pc) and full 
calculation of Gashi et al.\cite{gashi1} is very small except for the s 
wave. The dotted curve represents a fit to the NORDITA 
results\cite{nordita} to be used later and given in the appendix.} 
\label{ccplus} \end{figure}
\end{center}

\begin{center}
\begin{figure}[htb]
\epsfig{file=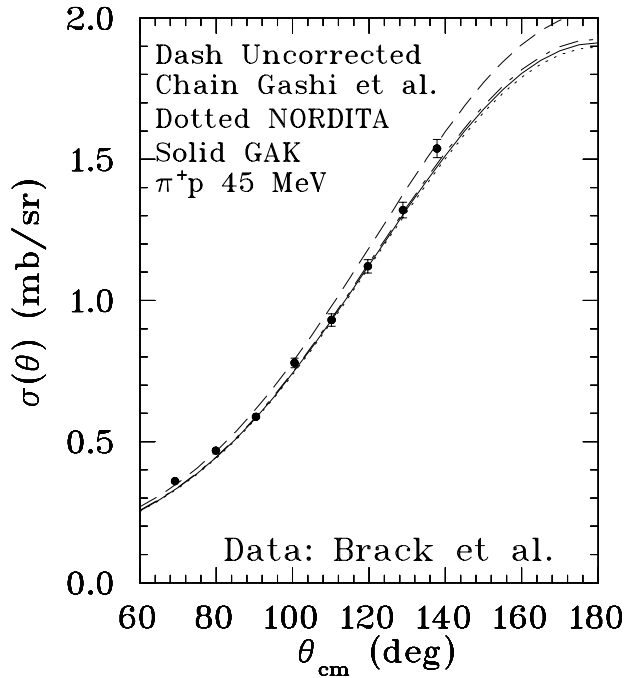,height=3.5in} 
\caption{Effect of the corrections for $\pi^+$p scattering in a typical 
case.  It is seen that the corrections are small but not negligible.
The three methods compared give very nearly the same result. The data
are from Brack et al. \cite{brack}.}
\label{wwo} \end{figure}
\end{center}

This phase shift is then compared with the hadronic phase shift,
$\delta_{\ell\pm}^h$, i.e. the one that would exist if there were no
Coulomb interaction.  A potential model is used to calculate these two
quantities in the present case.  In the limit that the
interactions are very weak, the hadronic and nuclear phase shifts become
equal so that the difference may be expected to be small.  The
correction is defined as the difference between them.

\eq
C_{\ell\pm}=\delta^n_{\ell\pm}-\delta^h_{\ell\pm}
\qe

Figure \ref{ccplus} shows a comparison of the results the different 
determinations. It is seen that the potential methods are in reasonable 
agreement. The point Coulomb result of Gashi et al.\cite{gashi1}
included no higher order electromagnetic effects whereas the full 
calculation included a finite charge distribution as well. Since the
corrections from GAK included a finite charge density but no no higher 
order electromagnetic effects, there is no exact comparison possible.
The dispersion relation approach is seen to give only slightly different 
results.

While one can judge to some extent the size and degree of agreement from
Fig. \ref{ccplus}, it is useful to see the effect on the cross section
itself. Figure \ref{wwo} shows such a comparison at a kinetic energy of 45
MeV. The hadronic phase shifts are taken for Ref. \cite{gak} for all of
the cases. It is seen that the correction is small but not negligible,
being of the order of two standard deviations of accurate data.  The
differences \underline{between} the corrections would seem to be
unimportant at the present level of the data.

\section{$\pi^-$ proton scattering and charge exchange}

In this case the expression of the corrections is more complicated since 
the potential
system consists of a pair of coupled differential equations. Not
only is there a correction due to the finite charge in the $\pi^-$p
channel but the neutron-proton and $\pi^0-\pi^{\pm}$ mass differences 
need to be taken into account.

\subsection{Method}

To express the corrections in the coupled channel case we follow the
formulation of Gashi et al.\cite{gashi2}.
The solution of the coupled channel system\cite{pete,gak} produces 
the
amplitude for $\pi^-$ proton scattering, charge exchange and $\pi^0$ 
neutron scattering. The result is expressed as  a 2 by 2 symmetric S 
matrix with the diagonal elements
representing $\pi^-$p and $\pi^0$n elastic scattering and the
off-diagonal elements charge exchange.  The scattering matrix is expressed
in terms of a 2 by 2 real K matrix.

\eq
S=\frac{1+iK}{1-iK};\ \ \ K=-i\frac{S-1}{S+1}.
\qe

These equations hold for each partial wave where we have suppressed their
indices. The $S_{11}$ matrix element has been multiplied by 
$e^{-2i\sigma_{\ell}}$ to remove the pure Coulomb effect before the 
calculation of the K matrix. Since the K-matrix is real and symmetric it 
can be transformed to a diagonal form with an orthogonal matrix.

\eq
\left(
\begin{array}{cc}
\cos\phi&-\sin\phi\\
\sin\phi&\cos\phi\\
\end{array}\right)
\left(
\begin{array}{cc}
K_{11}&K_{12}\\
K_{21}&K_{22}\\
\end{array}\right)
\left(
\begin{array}{cc}
\cos\phi&\sin\phi\\
-\sin\phi&\cos\phi\\
\end{array}\right)
=
\left(
\begin{array}{cc}
K_1&0\\
0&K_2\\
\end{array}\right)
\equiv
\left(
\begin{array}{cc}
\tan\delta_1&0\\
0&\tan\delta_2\\
\end{array}\right).
\label{trans}\qe

\begin{center}
\begin{figure}[htb]
\psfig{file=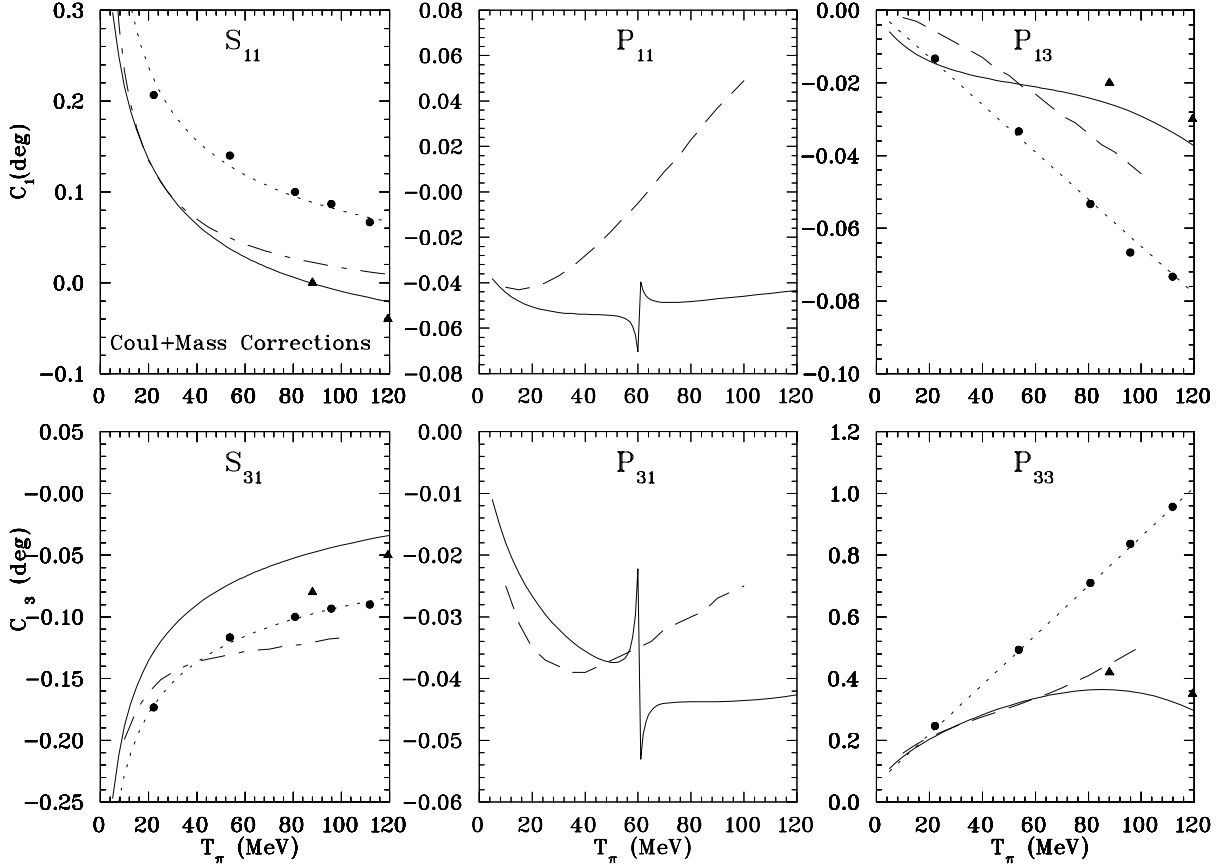,angle=90,height=4.5in} 
\caption{The solid curve gives the result with both Coulomb and mass 
difference corrections from GAK. The dashed line is from Gashi et 
al.\cite{gashi2}, the solid dots are from Tromborg et al.\cite{nordita} 
and the solid triangles are from Zimmermann\cite{zimmermann}. The dotted
lines represent a fit to the NORDITA\cite{nordita} results used later.}
\label{deltacm}
\end{figure}
\end{center}

\begin{center}
\begin{figure}[htb]
\epsfig{file=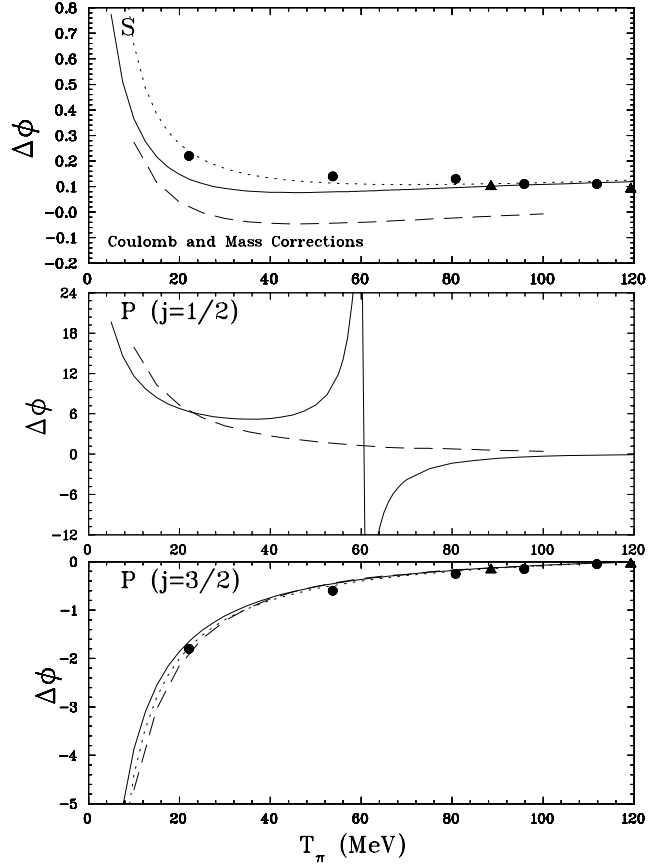,height=4.5in} 
\caption{Comparison with the correction for the rotation angle 
$\Delta\phi$. The meaning of the symbols is the same as in Fig. 
\protect\ref{deltacm}}
\label{phicm}
\end{figure}
\end{center}

If isospin is a good quantum number then
$\cos\phi\rightarrow\sqrt{\frac{2}{3}} \equiv \cos\phi_0$ and $\delta_1$
and $\delta_2$ are the isospin $\h$ and $\frac{3}{2}$ phase shifts.  The
form expressed in Eq. \ref{trans} is valid independent of isospin
conservation and the differences of the resultant phase shifts $\delta_1$,
$\delta_2$ as well as the value of $\phi$ needed to diagonalize the 
system, from the unbroken isospin values are used to quantify the 
breaking.

The corrections are expressed as
\eq
C_{1}\equiv \bar{\delta}_{\h}-\delta^h_{\h};\ \ 
C_{3}\equiv \bar{\delta}_{\frac{3}{2}}-\delta^h_{\frac{3}{2}};
\ \ \Delta\phi\equiv \bar{\phi}-\phi^h=\bar{\phi}-\phi_0
\qe
for each partial wave, where the barred quantities are those obtained 
from the model by solving for $\delta_1$, $\delta_2$ and $\phi$ from the
equations above with Coulomb and mass differences present.  The quantities 
with super script ``h'' (the hadronic values) are those obtained from the 
solution of the system with all pion masses equal to the charged pion 
mass, the neutron mass equal to the proton mass and no Coulomb interaction. 

\begin{center}
\begin{figure}[htb]
\psfig{file=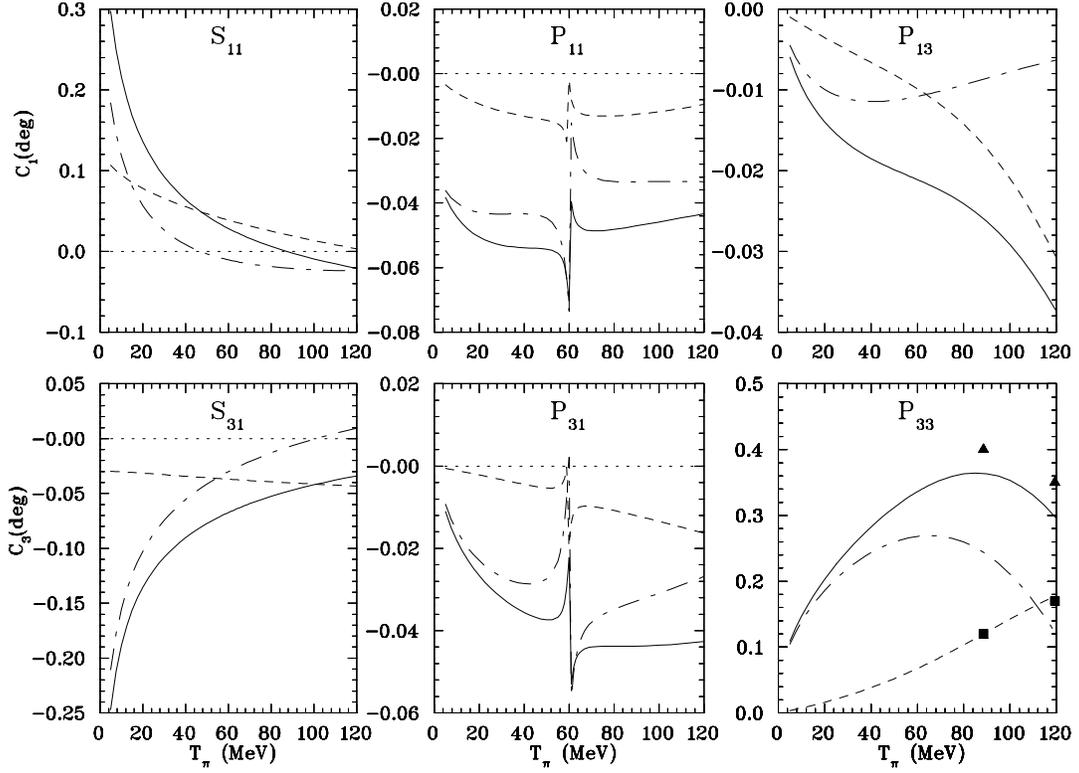,angle=90,height=4.in} 
\caption{The short dashed line contains the Coulomb correction only,
the dash-dot line is with the mass difference correction only and
the solid line contains both corrections. The results of Zimmermann are 
given by the solid triangles (full calculation) and solid squares 
(Coulomb correction only).} \label{deltacmcm}
\end{figure}
\end{center}

\begin{center}
\begin{figure}[htb]
\epsfig{file=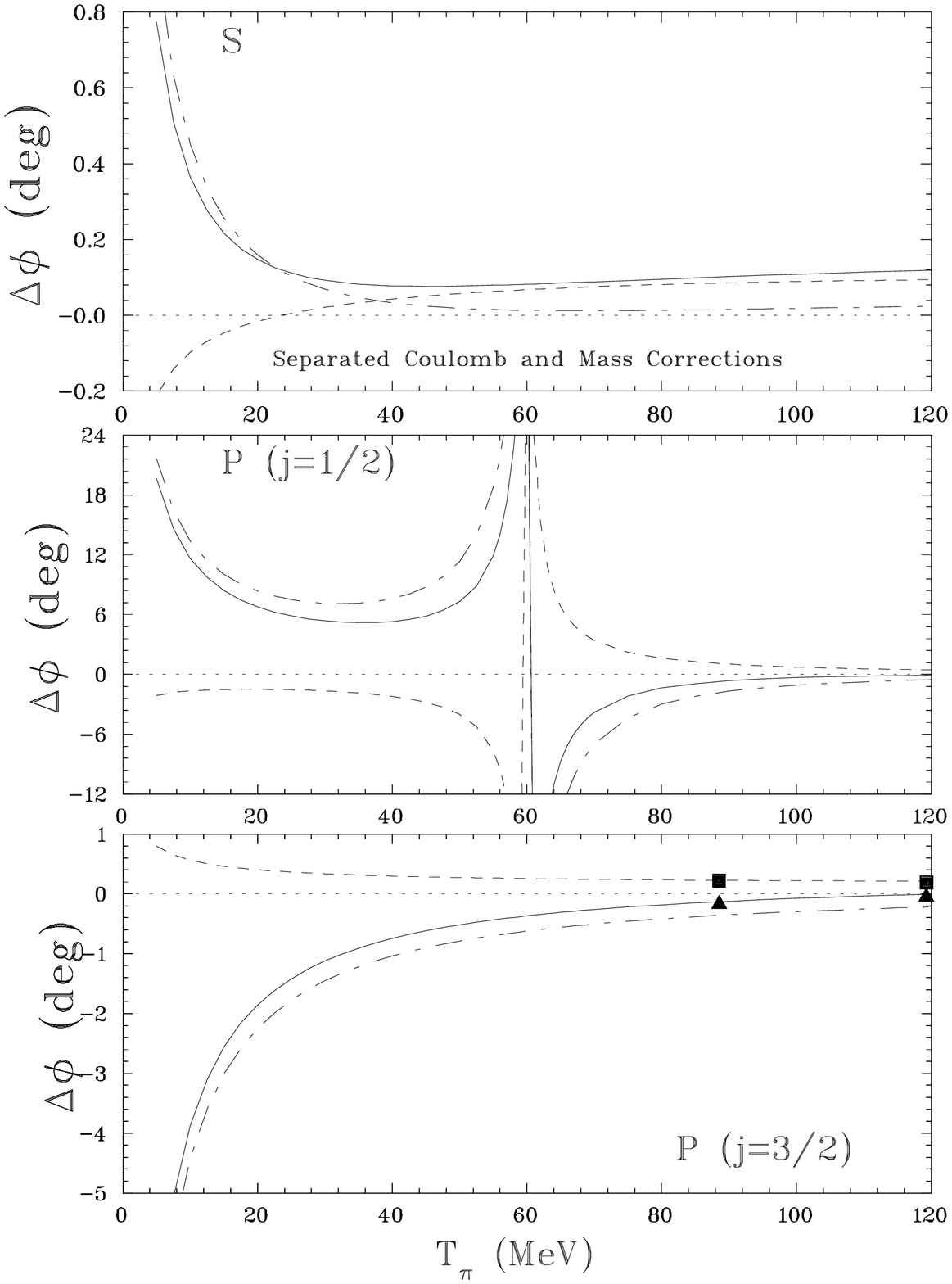,height=4in} 
\caption{Individual contributions to $\Delta\phi$. The convention for the 
lines and points is the same as in Fig. 
\protect\ref{deltacmcm}}
\label{phicmcm}
\end{figure}
\end{center}

The hope is that these corrections will be nearly model independent and
one can fit the data by choosing $\delta_1^h$ and $\delta_2^h$, apply the
corrections to obtain the full $K$ and $S$ matrices, add in the Coulomb
amplitude where appropriate and compare with data to calculate a value of
$\chi^2$. A search on the values of the isospin pure phase shifts,
$\delta_1^h$ and $\delta_2^h$ can then be made.

We now address the problem of calculating the corrections using the 
model. Since $K_1$ and $K_2$ are the eigenvalues of the K-matrix we can 
solve directly for them with
\eq
\lambda=\frac{K_{11}+K_{22}\pm\sqrt{(K_{11}-K_{22})^2+4K_{12}^2}}{2}.
\label{eigen}\qe
We may now identify the plus and minus signs in this expression 
with the isospin $\h$ and $\frac{3}{2}$ states.

Multiplying Eq. \ref{trans} from the left by the transpose of the rotation 
matrix we
find four relations for the $\tan\phi$, all of which are equivalent.
Two of them are:
\eq
\tan\phi=\frac{K_{11}-K_1}{K_{12}};\ \ 
\tan\phi=\frac{K_{12}}{K_2-K_{11}}.
\qe

\begin{center}
\begin{figure}[htb]
\epsfig{file=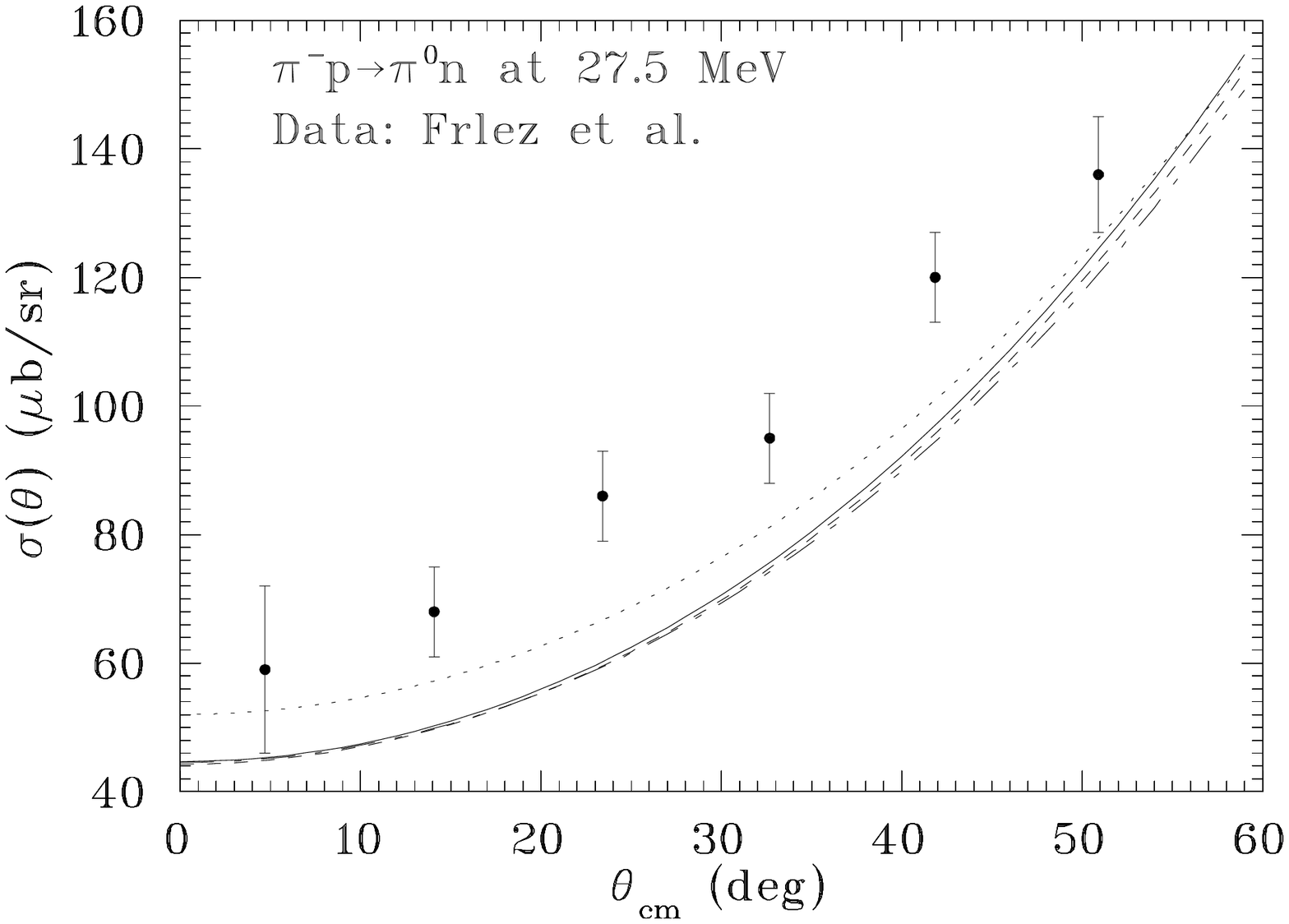,angle=90,height=3in} 
\caption{Comparison of methods of corrections for charge exchange for 
forward angles at low energies with the data of Frlez et al.\cite{frlez}. 
The dotted curve represents no correction, the solid curve gives the 
correction from GAK\cite{gak}, the chain-dash curve is from Gashi et 
al.\protect\cite{gashi2} and the dashed curve represents the corrections 
from NORDITA\protect\cite{nordita}}
\label{comparecex}
\end{figure}
\end{center}

\begin{center}
\begin{figure}[htb]
\epsfig{file=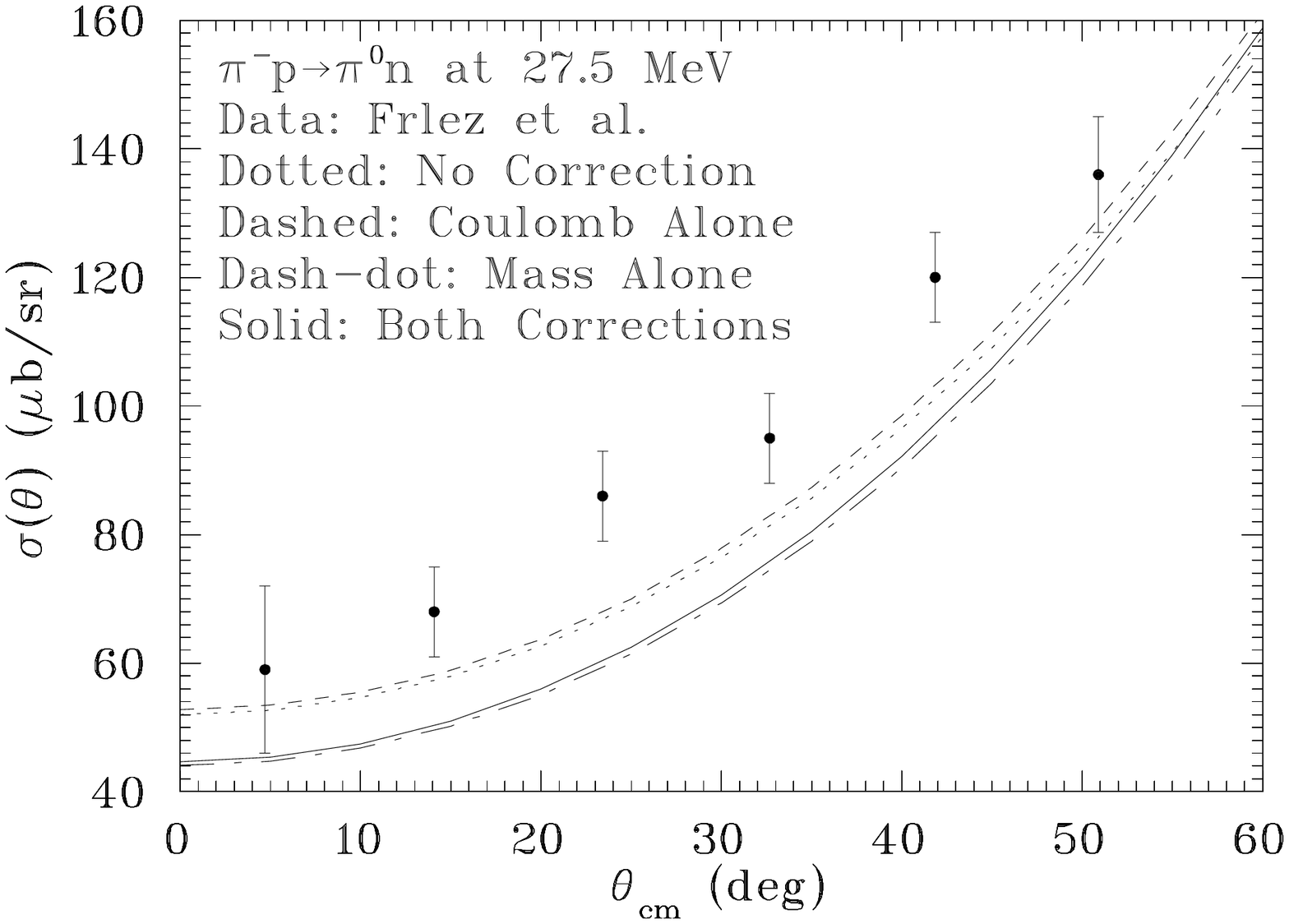,angle=90,height=3in} 
\caption{Separation of the Coulomb and mass corrections Compared with 
the data of Frlez et al.\cite{frlez}}
\label{frlez}
\end{figure}
\end{center}

\subsection{Difficulty with the P$_{\h}$ wave}

While in most cases there is no difficulty in choosing the proper sign in
Eq. \ref{eigen} to make the correct association of $\delta_1$ and
$\delta_2$ with the isospin phase shifts, in the case of the P$_{\h}$
scattering a problem arises.  In this case around 60 MeV
the two hadronic phase shifts (P$_{31}$ and P$_{11}$) cross.  For the 
two eigenvalues, $K_1$ and $K_2$ to be equal the radical in Eq. 
\ref{eigen} must vanish. This requires that the conditions
\eq
K_{11}=K_{22};\ \ K_{12}=0\label{both}
\qe
be satisfied simultaneously.  For unbroken isospin this indeed happens.
In this case isospin requires
\eq
K_{12}=\sqrt{2}(K_{22}-K_{11})
\qe
so that
\eq
\lambda=\frac{K_{11}+K_{22}\pm 3\sqrt{(K_{11}-K_{22})^2}}{2}=
\frac{K_{11}+K_{22}\pm 3|K_{11}-K_{22}|}{2}
\qe

In order for $\lambda$ to have a continuous derivative the $\pm$ 
association must be changed at the zero of the radical.  If one
chooses the sign such that the radical has the actual
sign of $K_{11}-K_{22}$ then
\eq
\lambda_+=2K_{11}-K_{22}=\tan\delta^{\frac{3}{2}};\ \ 
\lambda_-=2K_{22}-K_{11}=\tan\delta^{\frac{1}{2}}.
\qe

For broken isospin, in general, both conditions in Eq. \ref{both} will NOT 
be satisfied at once so that the two eigenvalues can never be equal.  
To understand what happens in this case it is useful to look at the direct 
solution of Eq. \ref{trans}.

In the general case we can expand Eq. \ref{trans} to find that the 
condition for the off-diagonal elements to be zero is
\eq
2\phi=\tan^{-1}\frac{2K_{12}}{K_{22}-K_{11}}
\qe
so that $2\phi$ changes by $\pi$ around the point $K_{22}=K_{11}$.
Consider a simple example in which the isospin relation is slightly 
broken in a simple way such that
\eq
K_{12}=\sqrt{2}(K_{22}-K_{11})+\epsilon
\qe
then
\eq
2\phi=\tan^{-1}\left(2\sqrt{2}+\frac{2\epsilon}{K_{22}-K_{11}}
\right)
\qe
so that $2\phi$ increases by $\pi$ as the difference $K_{22}-K_{11}$ 
changes sign. We have assumed that $K_{22}>K_{11}$
for energies lower than the crossing point.  If the opposite is true, a
very similar argument gives the same effect with 2$\phi$ decreasing by 
$\pi$.

The values of $K_1$ and $K_2$ are given by
\eq
K_1=\h K_{11}(1+\cos 2\phi)+\h K_{22} (1-\cos 2\phi)-K_{12}\sin 2\phi
\stackrel{\phi\rightarrow\phi_0}
{\rightarrow}
\tan\delta^{\frac{3}{2}}\qe

\eq
K_2=\h K_{11}(1-\cos 2\phi)+\h K_{22} (1+\cos 2\phi)+K_{12}\sin 2\phi
\stackrel{\phi\rightarrow\phi_0}{\rightarrow}
\tan\delta^{\frac{1}{2}}\qe
Indeed, as $2\phi$ increases by $\pi$, $K_1\leftrightarrow K_2$
since the sine and cosine change sign under these conditions.

Thus, the correction is non-perturbative. The change in $2\phi$ by $\pi$
occurs at the point where the hadronic phase shifts cross and hence 
depends strongly on the hadronic phase shifts used to define the 
corrections. In this case the general method fails except for corrections 
far from the crossing point.

 One can choose to change the association of the plus and minus signs at
the zero of $K_{12}$ (that is what is done in the present work) or when
$K_{11}=K_{22}$ but either way there will be discontinuities in the
corrections extracted at the crossing point.  The actual amplitudes remain
continuous, it is only this representation of the breaking which shows an
anomalous behavior but the compensation only occurs for the hadronic phase
shifts used in the calculation of the correction. If one uses these
corrections with phase shifts other than the ones for which the
corrections were derived, discontinuities in the resulting amplitudes will
result.  Of course, this is a small correction in a small partial wave in
this case so in a practical sense the problem may not be very serious.

The difficulty comes partly from the fact that, even for good isospin, for
$K_{1}=K_{2}$ any value of $\phi$ is valid for the transformation so
$\phi$ is undetermined at that point.  For unbroken isospin there is no
problem, however, since $\phi$ is determined arbitrarily close to the
crossing point on either side.

\begin{center}
\begin{figure}[htb]
\epsfig{file=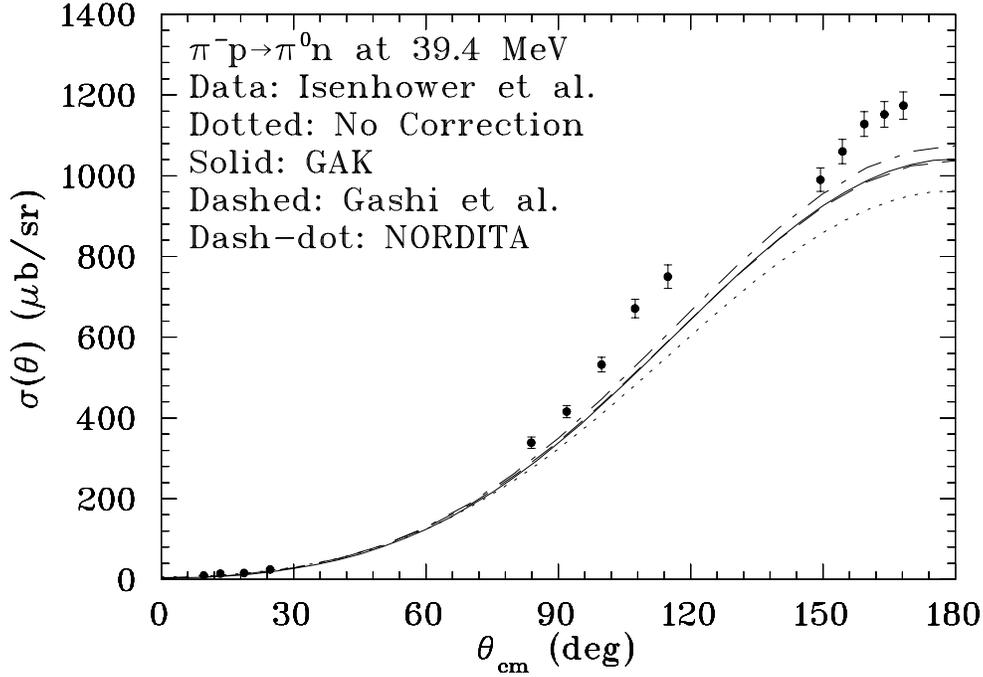,angle=90,height=3.5in} 
\caption{Comparison of the effects of the corrections 
with the charge exchange data of Isenhower et al.\cite{isen} at 39.4 
MeV. Other comparisons at 10.6 and 20.6 MeV are similar.}
\label{isen40}
\end{figure}
\end{center}

\begin{center}
\begin{figure}[htb]
\epsfig{file=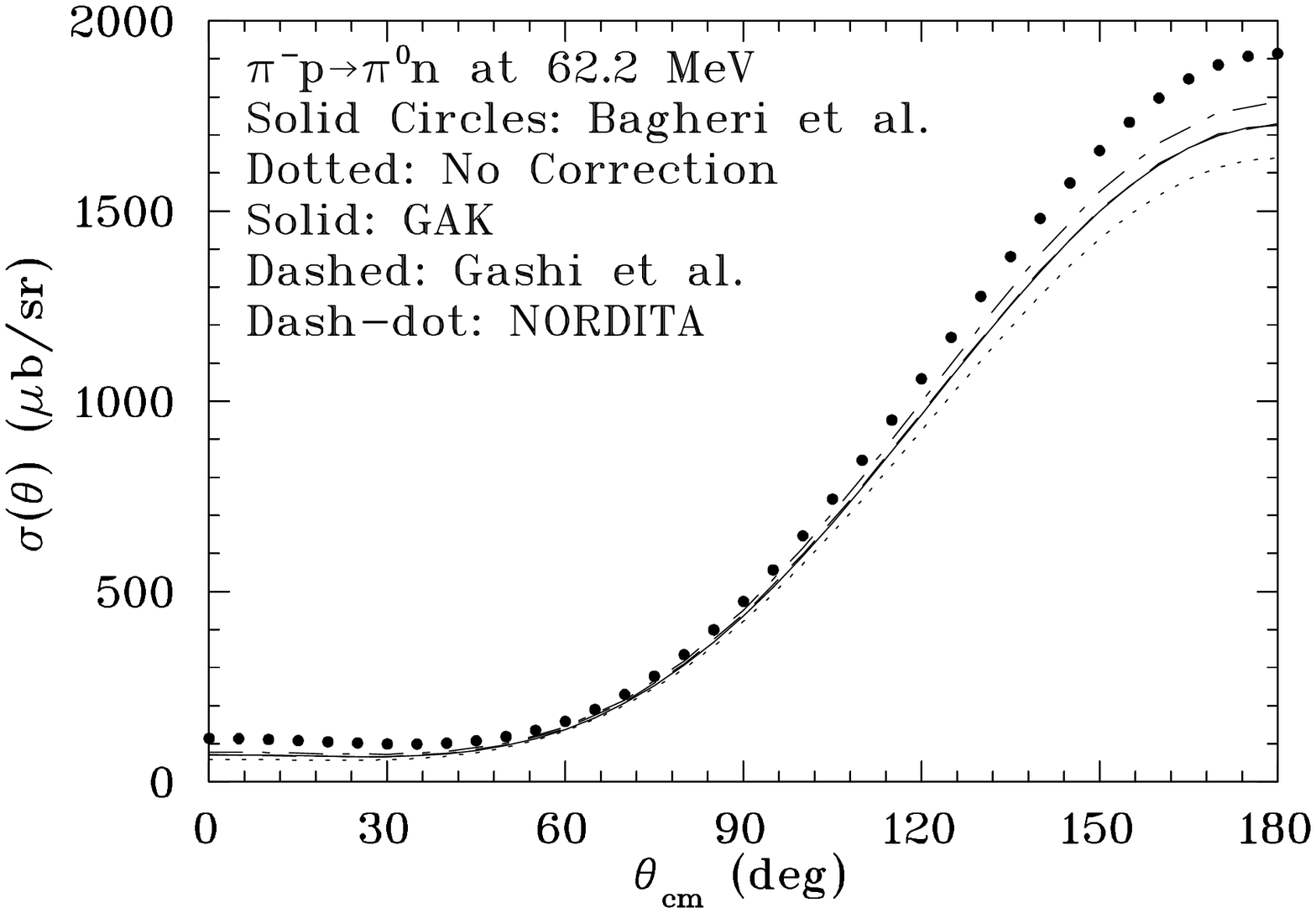,angle=90,height=3.5in} 
\caption{Comparison of the effect of the corrections with the data of 
Bagheri et al.\protect\cite{bagheri} at 62.2 MeV. Other comparisons with 
data from the same group at 45 and 92 MeV are similar.}
\label{measday62}
\end{figure}
\end{center}

\subsection{Comparison with previous models}

An obvious isospin breaking effect due to the mass differences leads to a
positive Q-value for $\pi^-p\rightarrow \pi^0n$ and a negative Q-value for
$\pi^0n\rightarrow \pi^-p$. This fact destroys the symmetry of the
amplitudes derived from the symmetric S-matrix, due to a factor $k_f/k_i$
originating from the incoming and outgoing fluxes which is obtained in the
charge exchange cross section obtained from the procedure above. This
factor is included in the calculations of charge exchange cross sections
to be shown in this section.

Figures \ref{deltacm} and \ref{phicm} show a comparison of the corrections
obtained from GAK\cite{gak} with previous work. It is seen that there is
qualitative agreement among all calculations.  There is rather good
agreement between the present work and Zimmermann at the higher energies
where he calculated. For the S$_{11}$ and P$_{33}$ partial waves there is
good agreement with the work of Gashi et al.\cite{gashi2} as for the case
of $\Delta\phi$ for the P$_{\frac{3}{2}}$ channel where, in fact, all
calculations agree.

Figures \ref{deltacmcm} and \ref{phicmcm} give the breakdown of the
contributions to the corrections into Coulomb and mass differences for
GAK. It is seen that the most important contribution is from the mass
differences, especially at low energies.  Since Zimmermann gave a
separation of the two effects for the P$_{33}$ phase shift and the
$\Delta\phi$ for the P$_{\frac{3}{2}}$ channel a comparison can be made 
and the agreement is good. The NORDITA\cite{norditab} group also
gave a separation into the Coulomb and mass difference contributions. The
mass difference correction dominates at low energy (see Ref.  
\cite{norditab}, Figures 2, 3 and 4).  Hence, while these corrections are
often called electromagnetic, it is the correction arising from mass
differences which is the most important.

While the expression of the correction in this form is very useful, it is
difficult to appreciate the effect on the cross sections of the
differences among the determinations as they may well be compensating to
give similar corrections to physical observables. Hence, we consider the
effects of these corrections on some charge exchange cross sections and
the $\pi^-$p elastic amplitude.  In each case the ``no correction''
calculation is made from the hadronic phase shifts of GAK\cite{gak} and
the corrections are made relative to them.

Figure \ref{comparecex} shows a comparison of the correction with the
charge exchange data of Frlez et al.\cite{frlez} at 27.5 MeV for the three
groups who calculated in this energy range.  It is seen that the resulting
corrected cross sections are all very nearly the same and that the
correction takes the result away from the data whereas the
uncorrected calculation (dotted line) lies much closer to agreement.
This last effect occurs only at forward angles.

Figure \ref{frlez} gives the breakdown of the contributions. It is
seen that the Coulomb correction actually moves the result slightly closer 
to the data while the mass difference correction takes it away.

Figure \ref{isen40} shows the effect over the full angular range compared
with the data of Isenhower et al.\cite{isen}. It is seen that for larger
angles the corrections move the prediction closer to the data.

At higher energies the effect of the corrections is somewhat smaller. A
comparison with the data of Bagheri et al.\cite{bagheri} is shown in Fig.
\ref{measday62}.  The experimental results were given as coefficients of
Legendre polynomials in this case and the dots shown were calculated from
these numbers. The effect of the corrections is seen to be small at this
energy and the two potential models agree well.

In general, the difference between prediction and data can be expressed as
a difference in normalization of around 15\%.

\begin{center}
\begin{figure}[htb]
\epsfig{file=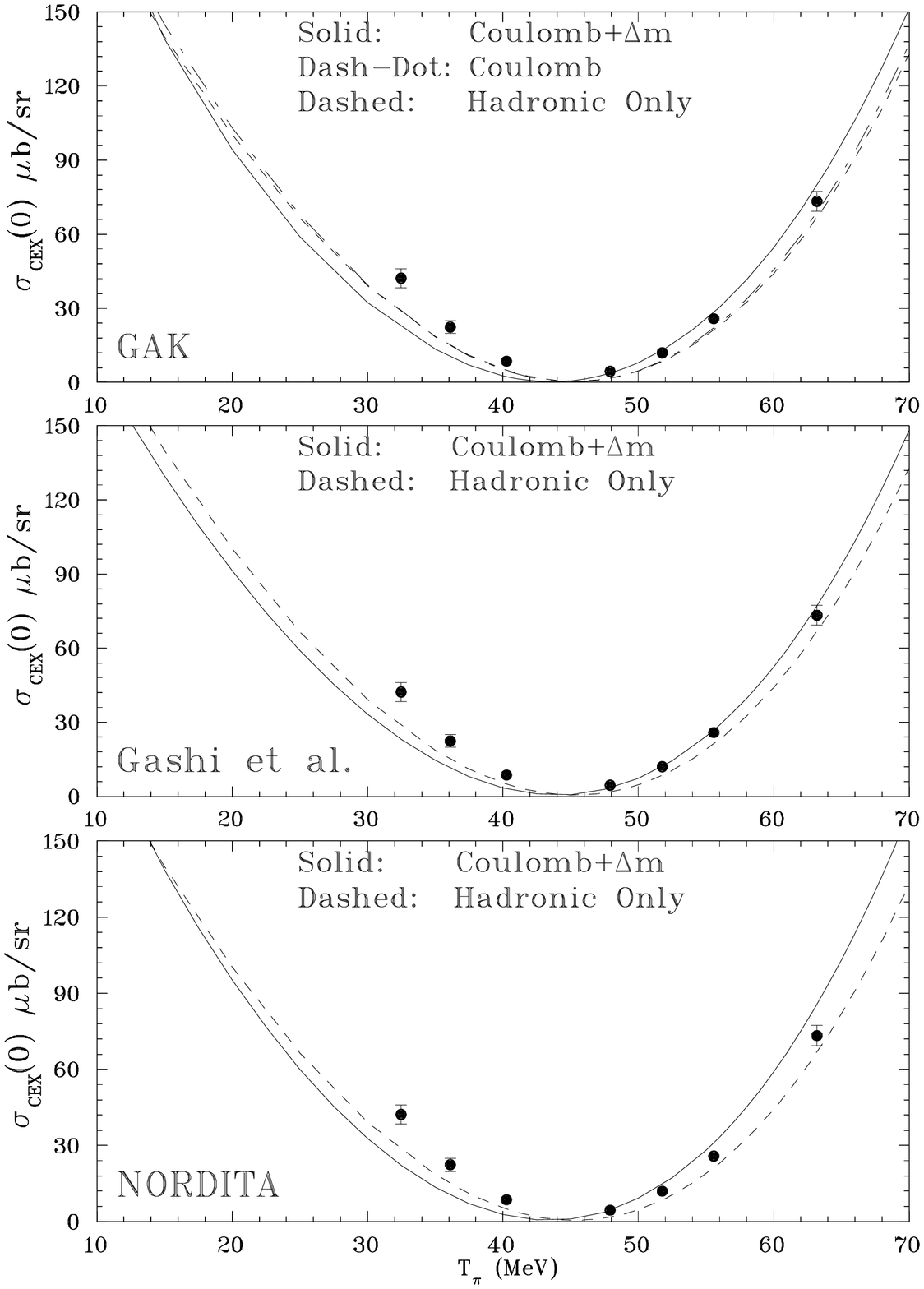,height=4.5in}
\epsfig{file=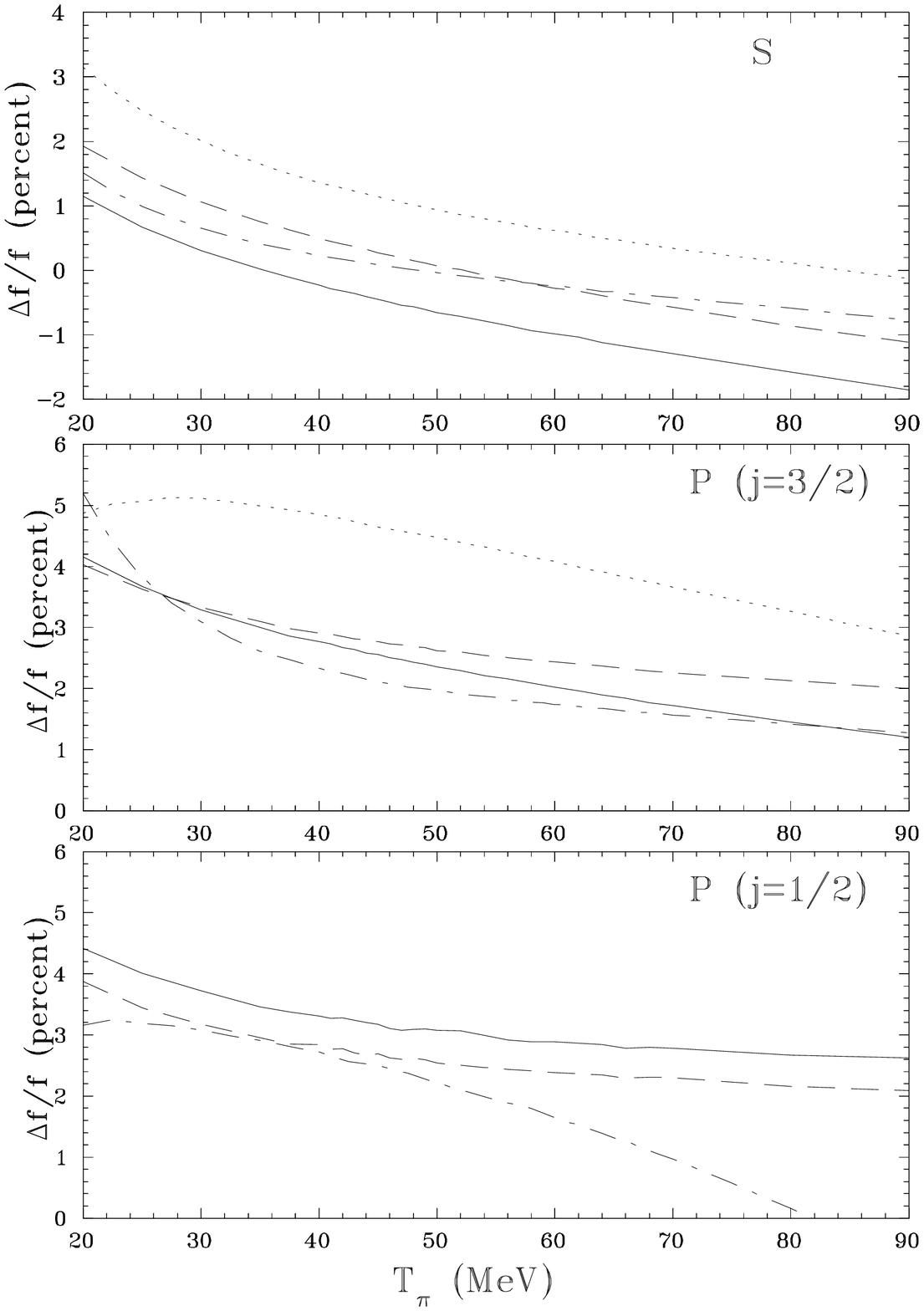,height=4.5in} 
\caption{Left: Zero degree charge exchange with corrections from 
GAK\cite{gak}, Gashi et al.\cite{gashi2} and NORDITA\cite{nordita}.
Right: Corrections to the real part of $f_{\pi^-p\rightarrow\pi^-p}$
for GAK\cite{gak} (dash: Coulomb only, solid: full correction),
Gashi et al.\cite{gashi2} (dash-dot) and NORDITA\cite{nordita}
 (dotted).}
\label{forward}
\end{figure}
\end{center}

The forward charge exchange cross section is a sensitive measure of
isospin breaking since it shows a deep minimum near 45 MeV.  While most
checks on isospin depend on obtaining an absolute cross section, the 
position of this minimum does not; only a knowledge of the beam momentum 
is needed.

The left part of Fig. \ref{forward} shows the effect of the corrections of
the three groups compared with the data of Fitzgerald et
al.\cite{fitzgerald}. We see that the corrections are all very nearly the
same and move the predicted cross section away from the data.  While the
uncorrected prediction cannot be said to give a good fit to the data, it
is much better than the prediction after correction for the mass
differences. The Coulomb potential has very little effect on the position
of the minimum.

The right part of Figure \ref{forward} shows the prediction of the
corrections for the amplitude for $\pi^-p\rightarrow\pi^-p$. It is seen
that the mass correction is significantly different from the pure Coulomb
effect. One can also note a large difference among the methods.

\subsection{Comparison with Chiral Perturbation Theory}

While the results of the previous section are rather consistent, at least
for the charge exchange, they may all be wrong.  Thus, it is very
desirable to have a comparison with an independent method calculating from
an approximate QCD viewpoint.

A calculation of isospin breaking in Chiral Perturbation Theory (ChPT) has 
been made by Fettes and Mei\ss ner\cite{fettes}.  They calculated 6 ratios 
which are measures of isospin breaking.  Most of the ratios (3-6) involve
quantities difficult to measure. Ratio number one involves isoscalar
quantities which vanish at threshold so that the ratio will depend
sensitively on the phase shift fit used to compute them.

Their second ratio, 
\eq 
R_2= 2\frac{f_{\pi^+p\rightarrow\pi^+p}-f_{\pi^-p\rightarrow\pi^-p}
-\sqrt{2}f_{\pi^-p\rightarrow\pi^0n}}
{f_{\pi^+p\rightarrow\pi^+p}-f_{\pi^-p\rightarrow\pi^-p}
+\sqrt{2}f_{\pi^-p\rightarrow\pi^0n}} 
\qe 
involves the same expression as is used to test isospin in the charge 
exchange reaction so is very suitable for comparison.  As stated by the 
authors, a direct comparison can not be made between their work and data 
but one can calculate the ratio with the hadronic model under the same 
conditions.

If one were to attempt to compare with actual charge exchange amplitudes 
obtained directly from the data they would include the flux factor
 $\sqrt{k_f/k_i}$. Since the ChPT calculation is made at the matrix 
element level, this factor should not be in the comparison.  We show
both cases to demonstrate the importance of the factor.

Since they calculated {\em corrections} to the strong process their 
combination
\eq
f_{\pi^+p\rightarrow\pi^+p}-f_{\pi^-p\rightarrow\pi^-p}
\qe
goes to zero in the limit of no strong interaction and we must remove
the pure Coulomb amplitude. We might subtract it partial wave by
partial wave or simply use the nuclear phase shifts.
It is the latter prescription which is shown in the comparison
graphs.  There is, in fact, very little difference in these two
ways of removing the Coulomb effect.

\begin{center}
\begin{figure}[htb]
\epsfig{file=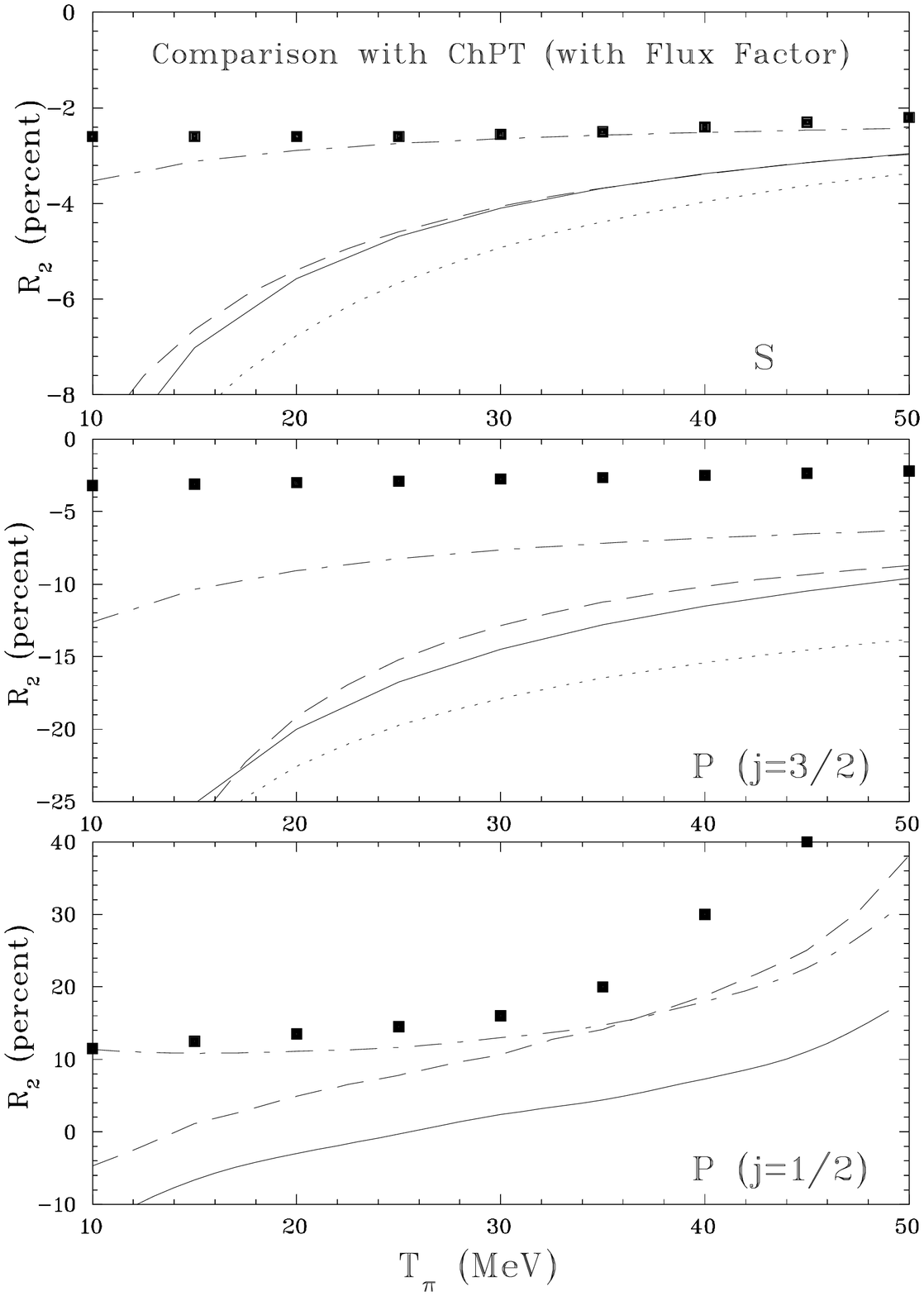,height=4.5in} 
\epsfig{file=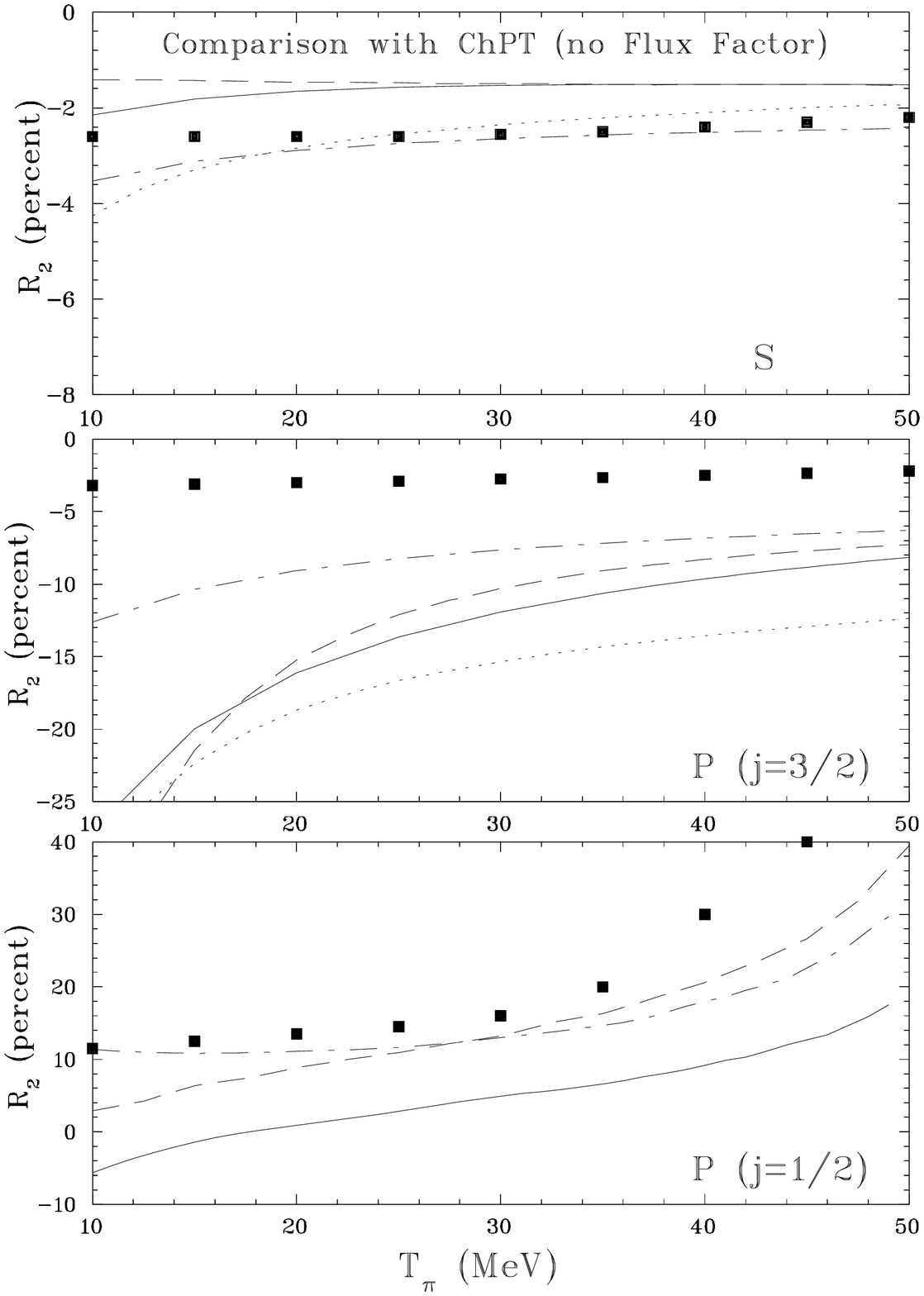,height=4.5in} 
\caption{The left figure shows the ratio $R_2$ calculated with GAK 
corrections for Coulomb only (dash-dot) and Coulomb plus mass differences 
(solid) compared with that from Fettes and Mei\ss ner\cite{fettes} (solid 
squares). The dashed line gives the result with the corrections of Gashi 
et al.\cite{gashi2} and the dotted line gives that of 
NORDITA\cite{nordita}. The flux factor $\sqrt{k_f/k_i}$ is included in the 
hadronic calculations. The right figure shows the same information except 
that the flux factor $\sqrt{k_f/k_i}$ is not included in the hadronic 
calculations.}
\label{fettescm}
\end{figure}
\end{center}

Figure \ref{fettescm} shows the comparison with and without the flux
factor.  We see that, for the S and P$_{\h}$ waves the pure Coulomb
correction agrees rather well with the ChPT (full)  calculations. The
P$_{\h}$ ratio is large and shows a singularity around 60 MeV due to
crossing of the hadronic phase shifts leading to the vanishing of the
charge exchange amplitude in this partial wave.  The position of this
crossing depends on the detailed fit of the hadronic phase shifts and
hence cannot be expected to be the same in the different cases. For the
P$_{\frac{3}{2}}$ wave, even the ratio calculated with Coulomb corrections
alone is more than a factor of 2 larger than the ChPT result, perhaps due
to the fact that the amplitude is larger and the perturbation series has
not converged.

Without the flux factor, only the P$_\frac{3}{2}$ partial wave shows a
significant discrepancy between the models and the ChPT calculations. The
two potential models and the dispersion relation approach show about the
same factor of 3--5 difference in this case.  Since this is the channel in
which the Delta resonance occurs, one can question whether or not the
explicit degrees of freedom of the Delta can make an important difference.

\section{Conclusions}

We have seen that the hadronicly calculated corrections of GAK\cite{gak}
are in rough agreement with previous determinations. In most cases, the
values of GAK and Gashi et al.\cite{gashi1,gashi2} are in reasonable
agreement.  All of these methods give a significant (and very similar)
correction to the charge exchange cross section at low energies due to the 
mass differences.

Corrections calculated from ChPT are significantly smaller than those of 
the hadronic methods in the P$_{\frac{3}{2}}$ partial wave, especially 
when the mass correction is included. This may indicate that there is a 
general error in the hadronic corrections or that the Delta degree of 
freedom needs to be treated differently in the ChPT calculation.

Since the mass correction in the form currently used has been questioned
it is possible that the assumption about the isospin potentials remaining
unchanged is incorrect.  

If one assumes that the hadronic corrections are too large then a closer 
agreement can be achieved in the case of forward angle charge exchange
at low energy and the position of the forward minimum.  However, this 
conjecture does not correct the discrepancy at larger angles. So even
this rather drastic possibility does not solve the problem.

We thank G. C. Oades and Ulf Meissner for comments on the manuscript.

This work was supported by the National Science Foundation under contract
PHY-0099729.

\begin{appendix}
\section{Fits}

Following are a set of fits to the phase shifts and corrections which are
given as an aide. The form of the fits has no physical significance and
should not be used outside the range 5-100 MeV. In what follows, $T$ is
the laboratory kinetic energy in MeV and $\delta$ is in degrees.

\subsection{GAK Hadronic Phase Shifts}

The fit to the GAK hadronic phase shifts for the S waves is in the form
\eq
\delta=\frac{a\sqrt{T}}{1+bT+cT^2}.
\qe

\eq {\rm For\ }S_{1}\ \ a=1.0319;\ \ b=2.865\times 10^{-3};\ \ 
c=-8.0\times 10^{-7}
\qe
 
\eq {\rm For\ }S_{3}\ \ a=-0.5197;\ \ b=-5.37\times 10^{-3};\ \ 
c=1.939\times 10^{-5}
\qe
 
For the p-waves

\eq
\delta_{13}=-\frac{0.001806 T^{\frac{3}{2}}(1-0.00414 T)}
{1+0.001 T};\ \ 
\delta_{33}=\frac{0.01084T^{\frac{3}{2}}(1+0.0085 T)(1+.06e^{-0.057 T})}
{1-0.000436 T}
\qe

\eq
\delta_{11}=-\frac{0.00636T^{\frac{3}{2}}(1-0.006190 T-0.0304T^2)}
{1+0.0131 T};\ \ 
\delta_{31}=-\frac{0.00196T^{\frac{3}{2}}(1+0.00228 T)}{1+0.00011 T}
\qe

\subsection{$\pi^+$p Corrections}

The fits to the $\pi^+$p corrections by GAK are
\eq
C_{0+}=0.1+0.000049T;\ \ C_{1-}=-0.009+0.00045T;\  
C_{1+}=-0.0185-0.000397T^{1.5}
\qe

The fits to the $\pi^+$p corrections by Gashi et al.\cite{gashi1} are

\eq
C_{0+}=0.075+0.0005T;\ \ C_{1-}=0.005+0.0002T;\ 
C_{1+}=-0.033-0.0000195T^{2.2}
\qe
For NORDITA
\eq
C_{0+}=0.095;\ \ C_{1-}=0.00045T;\ \ C_{1+}=-0.035-0.000067T^2
\qe

\subsection{Coupled Channel Corrections}

The fit to the GAK corrections for the s wave is
\eq
C_1=0.753T^{-\h}(1-0.0114T);\ \ 
C_3=-\frac{0.619T^{-\h}(1-0.0051T)}{1.-0.00299T}
\qe
\eq
\Delta\phi=\frac{7.9}{T^{1.3}}+0.0011T-0.026
\qe
For the P$_{\h}$ partial wave
\eq
C_1=-0.035-\frac{.00103T}{1.+.0008T^2}+0.01745\tan^{-1}
\left(\frac{0.02}{T-60.1}\right)
\qe

\eq
C_3=0.019-\frac{0.01137T^{0.728}}{(1+0.137T^{0.72}+0.000039T^2)}-
0.01745\tan^{-1}\left(\frac{0.0155}{T-60.1}\right)
\qe

\eq
\Delta\phi=-1.2+\frac{148}{(T+4.3)^{0.95}}
-0.4363\tan^{-1}\left(\frac{2.0}{T-60.1}
\right)
\qe

For the P$_{\frac{3}{2}}$ partial wave

\eq
C_1=-0.001-0.0039T^{0.4}-0.0038\left(\frac{T}{100}\right)^4;\ \ 
C_3=0.1+0.0048T-0.23\left(\frac{T}{100}\right)^3;\ \ 
\Delta \phi=-\frac{44.5}{T}+0.37
\qe

The fit to the Gashi et al.\cite{gashi2} corrections used in this work is

\eq
{\rm For }\ s\ \        C_1=\frac{1.15}{T^{0.6}}-0.056;\ \ 
C_3=-\frac{0.3}{T^{0.3}}-0.04
\qe
\eq   
 {\rm For }\ p_{\frac{3}{2}}\ \   C_1=-0.0005T+0.006;\ \    
C_3=0.0036T+0.13
\qe
\eq
{\rm For }\ p_{\frac{1}{2}}\ \  
C_1=-0.045+0.000016T^{1.9};\ \ C_3=-0.05+0.00025T+0.046e^{-0.08T}
\qe

\eq
\Delta\phi_s=\frac{10.3}{T^{1.4}}+0.0013T-0.149;\ \
\Delta\phi_{p_{\frac{1}{2}}}=\frac{160}{T^{0.95}}-1.9;\ \ 
\Delta\phi_{p_{\frac{3}{2}}}=-\frac{45}{T-2}+0.4;  
\qe

The fit to the NORDITA\cite{nordita} corrections used in this work is

\eq
{\rm For }\ s\ \ C_1=\frac{1.1}{T^{0.4}}-0.095;\ \  
C_3=-\frac{0.55}{T^{0.25}}+0.082
\qe
\eq
{\rm For }\ p_{\frac{3}{2}}\ \ C_1=-0.00065T;\ \  C_3=0.008T+0.06
\qe
\eq
\Delta\phi_s=\frac{16}{T^{1.4}}+0.0008T+0.01;\ \ 
\Delta\phi_{p_{\frac{3}{2}}}=-\frac{48}{T}+0.4 
\qe

The NORDITA group did not give corrections for the P$_{\h}$ partial wave.

\end{appendix}

\end{document}